\documentstyle[12pt]{article}
\def \L{{\cal L}}
\def \delm{\partial_\mu}

\newcommand{\delum}{\partial^\mu}

\newcommand{\delmu}{\delum}

\def \mn{{\mu\nu}}
\newcommand{\nonum}{\nonumber\\}
\newlength{\mleng}
\newcommand{\mq}{\makebox[0.5\mleng]{ }}
\newcommand{\mqq}{\makebox[\mleng]{ }}
\newcommand{\mqqq}{\makebox[1.5\mleng]{ }}
\newcommand{\be}{\begin{equation}}
\newcommand{\ee}{\end{equation}}
\newcommand{\ba}{\begin{eqnarray}}
\newcommand{\ea}{\end{eqnarray}}
\def \tr{\hbox{tr}}
\newcommand{\Trang}[1]{\,\tr\left[{#1}\right]}

\begin{document}
\setlength{\baselineskip}{20pt}
\begin{titlepage}
\begin{flushright}
DPNU-92-43 \\
Sept. 1992
\end{flushright}
\begin{center}
{\huge Hidden Local Symmetry} \\
{\huge at One Loop}
\end{center}
\vfill
\begin{center}
{\large
Masayasu~{\sc Harada}%
\footnote{email:{\tt d42809a@nucc.cc.nagoya-u.ac.jp}}%
 \  and \  Koichi~{\sc Yamawaki}%
\footnote{email:{\tt b42060a@nucc.cc.nagoya-u.ac.jp}}%
\footnote{Work supported in part by the Takeda Science Foundation %
and the Ishida Foundation,
and also by the International Collaboration Program of the Japan Society
for the Promotion of Science.} %
}\\
\vspace{0.8cm}
{\it Department of Physics, Nagoya University} \\
{\it Nagoya 464-01, Japan}
\end{center}
\vfill
\begin{abstract}
We show that one-loop corrections with the hidden gauge boson loop
preserve in the low energy limit all the successful tree level predictions
of the hidden local symmetry
in the SU(2)${}_L$$\times$SU(2)${}_R$$/$SU(2)${}_V$ chiral Lagrangian.
Most amazingly, the $\rho$ meson dominance
of the pion electromagnetic form factor
survives the one-loop corrections at any momentum,
if and only if we take the parameter choice $a=2$.
For the choice $a=1$ (``vector limit"), $a$ is not renormalized ($Z_a=1$)
and no deviation from $a=1$ is induced
by the loop effects of the hidden gauge bosons.
Actually, $a=1$ is a nontrivial ultraviolet fixed point.
\end{abstract}
\end{titlepage}

It is now a popular notion to identify
the $\rho$ meson with the dynamical gauge boson
of the hidden local symmetry
in the SU(2)${}_L$$\times$SU(2)${}_R$/SU(2)${}_V$
nonlinear chiral Lagrangian\cite{BandoKugoYamawaki}.
By setting a parameter choice $a=2$ in this hidden local symmetry Lagrangian,
we can successfully reproduce three phenomenological facts%
\cite{BKUYY}:
The $\rho$-coupling universality, $g_{\rho\pi\pi}=g$,
the KSRF relation (II) ,
$m_\rho^2=2f_\pi^2g_{\rho\pi\pi}^2$\cite{KSRF}, and
the $\rho$ meson dominance of the electromagnetic form factor
of the pion\cite{Sakurai}.
Most remarkably, we obtain the celebrated KSRF (I) relation,
$g_\rho=2f_\pi^2g_{\rho\pi\pi}$, as an $a$-independent relation
which is actually characteristic to the hidden local symmetry and hence
may be regarded as a ``low energy theorem" of the symmetry%
\cite{BandoKugoYamawakiII}.
In fact it was proved
to be a low energy theorem at tree level\cite{BandoKugoYamawakiIII}.

Moreover, this hidden local symmetry Lagrangian has been applied to the
strongly coupled Higgs model and/or the effective theory of the
technicolor including the techni-$\rho$ meson
(``BESS model")\cite{Casalbuoni}.

Recently, one-loop corrections of the pion loop have been extensively studied
in the chiral Lagrangian (chiral perturbation theory%
\cite{Weinberg,GasserLeutwyler}),
which succeeded in reproducing systematically the low energy hadron physics
slightly away from the low energy limit dictated by the chiral symmetry.
However, this approach appears to fail
in reproducing the higher energy region,
even considerably lower than the $\rho$ meson pole,
and needs explicit degree of a new field, the $\rho$ meson.
Amazing fact is that the finite part of the one-loop counter terms
in the chiral perturbation theory is saturated
by the tree level effects of the $\rho$ meson\cite{Eckeretal}.
Thus the effective Lagrangian including the $\rho$ meson,
if it yields successful tree level results,
should be a good starting point towards constructing
a true ``effective field theory" including the quantum corrections.

As such we take the hidden local symmetry Lagrangian mentioned above.
Actually, it was pointed out%
\cite{Eckeretal,Wakamatsu}
 that loop effects of the vector mesons are
 crucial to the $\pi^+-\pi^0$ mass difference\cite{Das}.
Our goal is thus to promote the hidden local symmetry Lagrangian into
an ``effective field theory" valid up to beyond the $\rho$ meson pole by
including the full quantum effects.

In this paper we investigate one-loop effects
of the hidden local gauge boson,
and as a first step see whether or not the above successful tree level results
survive the loop effects.
First, we show that the ``low energy theorem" remains intact
even in the existence of the loop effects of the $\rho$ meson.
Second, we show that the $a$-dependent results ($a=2$),
the $\rho$-coupling universality and the KSRF(II) relation,
are satisfied in the low energy limit.
Third, we show the $\rho$ meson dominance
of the pion electromagnetic form factor
still holds at one-loop level, if and only if we take $a=2$.
Finally, we show that for $a=1$
there is no renormalization effect on $a$ ($Z_a=1$)
and the deviation from $a=1$ (``vector limit"\cite{Georgi})
is not induced by the loop effects of the hidden local gauge bosons.
This corresponds to the fact
that $a=1$ is a nontrivial ultraviolet fixed point
of the $\beta$ function of $a$.

Let us start with the
[SU(2)${}_L$$\times$SU(2)${}_R$]${}_{global}$$\times$[SU(2)${}_V$]${}_{local}$
``linear" model\cite{BKUYY}.
We introduce two SU(2)-matrix valued variables, $\xi_L(x)$ and $\xi_R(x)$,
which transform as
\be
  \xi_{L,R}(x) \rightarrow \xi'_{L,R}(x) = h(x) \xi_{L,R}(x) g_{L,R}^{\dag} ,
\ee
where $h(x)\in{[\mbox{SU(2)}_V]}_{local}$
and $g_{L,R}\in{[\mbox{SU(2)}_{L,R}]}_{global}$.
These variables are parameterized as
\be
  \xi_{L,R}(x) \equiv e^{i\sigma(x)/f_\sigma} e^{\pm i\pi(x)/f_\pi}
  \mqq [ \pi(x) \equiv \pi^a(x)\tau^a/2 ] ,
  \mq [ \sigma(x) \equiv \sigma^a(x)\tau^a/2 ],
\ee
where $\pi$ and $\sigma$ are the pion and the ``compensator"
(would-be Nambu-Goldstone field)
to be ``absorbed" into the hidden gauge boson (the $\rho$ meson),
respectively, and $f_\pi$ and $f_\sigma$ are the corresponding decay constants
in the chiral symmetric limit.
The covariant derivatives are defined by
\ba
&{}&
  D_\mu\xi_L(x) \equiv \delm \xi_{L}(x) - i g V_\mu(x) \xi_L(x)
    + i \xi_{L}(x) e {\cal B}_\mu(x) \frac{\tau_3}{2} ,
\nonum
&{}&
  D_\mu\xi_R(x) \equiv \delm \xi_{R}(x) - i g V_\mu(x) \xi_R(x)
    + i \xi_{R}(x) e {\cal B}_\mu(x) \frac{\tau_3}{2} ,
\ea
where $g$ is the gauge coupling constant of the hidden local symmetry,
$V_\mu$ ($\equiv V_\mu^a\frac{\tau^a}{2}$) the hidden gauge boson field
(the $\rho$ meson),
and ${\cal B}_\mu$ denotes the photon field
gauging the U(1)${}_{em}$ part of the
[SU(2)${}_L$$\times$SU(2)${}_R$]${}_{global}$.

Thus we obtain the Lagrangian of the
[SU(2)${}_L$$\times$SU(2)${}_R$]${}_{global}$$\times$[SU(2)${}_V$]${}_{local}$
``linear" model,
with the [SU(2)${}_L$$\times$SU(2)${}_R$]${}_{global}$ being partly gauged
by the photon field:\cite{BKUYY}
\be
 \L = \L_A + a \L_V + \L_{kin}(V_\mu) ,
\label{HiddenLagrangian}
\ee
where $a$ is a constant,
$\L_{kin}(V_\mu)$ denotes the (possibly induced) kinetic term
of the hidden gauge boson\cite{BandoKugoYamawaki},
and $\L_A$ and $\L_V$ are given by
\ba
&{}&
 \L_A = f_\pi^2
   \Trang{
     \bigl( \hat{\alpha}_{\mu\perp} \bigr)^2
   } ,
\nonum
&{}&
 \L_V = f_\pi^2
   \Trang{
     \bigl( \hat{\alpha}_{\mu\parallel} \bigr)^2
   } ,
\ea
with $\hat{\alpha}_{\mu\perp}$ and $\hat{\alpha}_{\mu\parallel}$
being the covariantized Maurer-Cartan 1-forms\cite{BandoKugoYamawakiIII}
\ba
&{}&
 \hat{\alpha}_{\mu\perp}(x) \equiv
 \frac{
   D_\mu\xi_{L}(x) \cdot \xi_L^{\dag}(x)
   - D_\mu\xi_{R}(x) \cdot \xi_R^{\dag}(x)
 }
 {2i} ,
\\
&{}&
 \hat{\alpha}_{\mu\parallel}(x) \equiv
 \frac{
   D_\mu\xi_{L}(x) \cdot \xi_L^{\dag}(x)
   + D_\mu\xi_{R}(x) \cdot \xi_R^{\dag}(x)
 }
 {2i} .
\ea
Normalizing the kinetic term of $\sigma$,
we find\cite{BandoFujiwaraYamawaki}
\be
  f_\sigma^2 = a f_\pi^2 .
\label{relFsigFpi}
\ee

Now the Lagrangian Eq.(\ref{HiddenLagrangian}) gives in the unitary gauge,
$\xi_L^{\dag}=\xi_R$ ($\sigma=0$),
the following tree level relations for
the $\rho$ meson mass $m_\rho$,
the $\rho$-$\gamma$ transition strength $g_\rho$,
the $\rho\pi\pi$ coupling constant $g_{\rho\pi\pi}$
and the direct $\gamma\pi\pi$ coupling constant $g_{\gamma\pi\pi}$:
\ba
&{}&
 m_\rho^2 = a g^2 f_\pi^2 ,
\label{rhomass}\\
&{}&
 g_\rho = a g f_\pi^2 ,
\label{grho}\\
&{}&
 g_{\rho\pi\pi} = \frac{1}{2} a g ,
\label{grhopipi}\\
&{}&
 g_{\gamma\pi\pi} = \left({1-\frac{a}{2}}\right) e .
\label{ggammapipi}
\ea

For a parameter choice $a=2$,
the above results reproduce the outstanding phenomenological facts%
\cite{BandoKugoYamawaki}:
\renewcommand{\labelenumi}{(\theenumi)}
\begin{enumerate}
\item $g_{\rho\pi\pi} = g$ (universality of the $\rho$-couplings)%
\cite{Sakurai},
\item $m_\rho^2 = 2g_{\rho\pi\pi}^2 f_\pi^2$
(KSRF II) \cite{KSRF},
\item $g_{\gamma\pi\pi} = 0$
($\rho$ meson dominance of the electromagnetic form factor of the pion)
\cite{Sakurai}.
\end{enumerate}
Moreover, Eqs.(\ref{grho}) and (\ref{grhopipi}) lead to
the KSRF relation\cite{KSRF} (version I)
\be
  g_\rho = 2 f_\pi^2 g_{\rho\pi\pi} ,
\ee
which is {\it independent of the parameter $a$}
and hence is the decisive test
of the hidden local symmetry\cite{BandoKugoYamawakiII}.
Thus it was conjectured
to be a ``low energy theorem"
of the hidden local symmetry\cite{BandoKugoYamawakiII} and was then
proved at tree level\cite{BandoKugoYamawakiIII}.

Now, we consider the one-loop effects of the gauge boson
of the hidden local symmetry.
Let us introduce the gauge-fixing terms corresponding
to the hidden gauge boson.
We define for the hidden local symmetry
an $R_\xi$ gauge condition so as to cancel the quadratic
vector-scalar mixing%
\footnote{Here we write the form
in which the BRS transformation is transparent.
Instead of the $\xi_L$ and $\xi_R$, we can write this term using $\sigma$%
\cite{CveticKogerler}, which does not alter
the results in this paper unchanged.}:
\ba
&{}&
\L_{GF}(V) \equiv
  - \frac{1}{\alpha}
  \Trang{ {(\delm V_\mu)}^2 }
  + \frac{i}{2} a g f_\pi^2
     \Trang{\delm V_\mu ( \xi_L - \xi^{\dag}_L + \xi_R - \xi^{\dag}_R ) }
\nonum
&{}& \mqqq\mqq
  + \frac{1}{16} \alpha a^2 g^2 f_\pi^4
   \Biggl\{
      \Trang{ ( \xi_L - \xi^{\dag}_L + \xi_R - \xi^{\dag}_R )^2 }
\nonum
&{}& \mqqq\mqqq\mqqq
      - \frac{1}{2}
      \left({
        \Trang{ \xi_L - \xi^{\dag}_L + \xi_R - \xi^{\dag}_R }
      }\right)^2
   \Biggr\} .
\ea
We also add the ghost Lagrangian corresponding to the gauge fixing:
\be
\L_{FP} \equiv
 i \Trang{
     \bar{v}
     \left\{{
       2 \delmu D_\mu v
       + \frac{1}{2} g^2 \alpha f_\pi^2 a
         ( v \xi_L + \xi^{\dag}_L v + v \xi_R + \xi^{\dag}_R v )
     }\right\}
   } ,
\ee
where $v$ denotes the ghost field.
In the following calculation we choose the Landau gauge, $\alpha=0$%
\footnote{The tree level results
Eqs.(\ref{rhomass})--(\ref{ggammapipi})
also hold in the Landau gauge.}.
In this gauge the would-be Nambu-Goldstone bosons $\sigma$ are still massless,
no other vector-scalar interactions are created and the ghost field couples
only to the gauge fields.
Since we are interested in the strong interaction effect,
we consider the photon field as the external field
and do not consider its loop effect.

For canceling the lowest derivative divergent part
we redefine the normalization of the parameters
and the fields such that
\ba
 &{}& a = Z_a a_r , \mqq e = Z_e e_r , \mqq g = Z_g g_r ;
\nonum
 &{}& V_\mu = Z_V^{1/2} V_{r\mu} , \mqq \pi = Z_\pi^{1/2} \pi_r ,
      \mqq \sigma = Z_\sigma^{1/2} \sigma_r ;
\nonum
 &{}& f_\pi = Z_\pi^{1/2} f_{\pi r} , \mqq
      f_\sigma = Z_\sigma^{1/2} f_{\sigma r} .
\ea
In the following calculations we define
the $\rho$ meson mass parameter $m_\rho$ by
\be
 m_\rho^2 \equiv a_r g_r^2 f_{\pi r}^2 .
\ee
Hereafter, we denote the pion momentum as $k_\mu$ and $q_\mu$,
and the $\rho$ meson momentum as $p_\mu$.
Throughout this paper we set the pion momentum on the mass-shell, $k^2=q^2=0$.

The one-loop graphs contributing to the $\rho\pi\pi$ coupling are
shown in Fig. \ref{fig:rhopipi}.
These contributions are given by
\ba
&{}&
  \Gamma_{(a)}^{\rho\pi\pi} =
  i g_r \epsilon_{cab} (k-q)_\mu \frac{a_r^3}{8}
  \frac{g_r^2}{{(4\pi)}^2}
  F_{(a)} (p^2) ,
\nonum
&{}&
  \Gamma_{(b)}^{\rho\pi\pi} =
  i g_r \epsilon_{cab} (k-q)_\mu
  \frac{g_r^2}{{(4\pi)}^2}
  F_{(b)} (p^2) ,
\nonum
&{}&
  \Gamma_{(c)}^{\rho\pi\pi} =
  i g_r \epsilon_{cab} (k-q)_\mu \frac{a_r(3a_r-4)}{48}
  \frac{p^2}{{(4\pi f_{\pi r})}^2}
  \left[{
    \frac{1}{\bar{\epsilon}} - \ln (-p^2) + \frac{8}{3}
  }\right] ,
\nonum
&{}&
  \Gamma_{(d)}^{\rho\pi\pi} =
  - i g_r \epsilon_{cab} (k-q)_\mu \frac{1}{24}
  \frac{p^2}{{(4\pi f_{\pi r})}^2}
  \left[{
    \frac{1}{\bar{\epsilon}} - \ln (-p^2) + \frac{8}{3}
  }\right] ,
\nonum
&{}&
  \Gamma_{(e)}^{\rho\pi\pi} = 0 ,
\nonum
&{}&
  \Gamma_{(f)}^{\rho\pi\pi} =
  i g_r \epsilon_{cab} (k-q)_\mu \frac{a_r}{48}
  \frac{p^2}{{(4\pi f_{\pi r})}^2}
  \left[{
    \frac{1}{\bar{\epsilon}} - \ln (-p^2) + \frac{8}{3}
  }\right] ,
\nonum
&{}&
  \Gamma_{(g)}^{\rho\pi\pi} = \Gamma_{(h)}^{\rho\pi\pi} = 0 ,
\label{gamrhopipi}
\ea
where
\ba
&{}&
 \frac{1}{\bar{\epsilon}} \equiv \frac{2}{4-n} - \gamma + \ln(4\pi) ,
\nonum
&{}& \mqqq
 [\gamma : \mbox{Euler constant},
 \mq n : \mbox{the dimension of the integral} ]
\ea
and $F_{(a)}(p^2)$ and $F_{(b)}(p^2)$ denote certain complicated functions
which have no divergent part and $F_{(a)}(p^2=0)=F_{(b)}(p^2=0)=0$.

{}From Eq.(\ref{gamrhopipi}) we can easily see that
at $p^2=0$ there exist
{\it no contributions} from the one-loop diagrams
and hence no counter terms;
\be
  Z_V^{1/2} Z_a Z_g Z_\pi - 1 = 0 .
\label{Zfactorrhopipi}
\ee
Then we find that the $\rho\pi\pi$ coupling
remains the same as the tree level
in the low energy limit;
\be
  g_{\rho\pi\pi}(p^2=0,k^2=0,q^2=0)
  = \frac{a_r}{2} g_r .
\label{couplinguniversality}
\ee
Eq.(\ref{couplinguniversality}) implies
that for $a_r=2$,
the universality of the $\rho$-couplings remains intact in the
low energy limit.

Similarly, one-loop graphs contributing to the $\rho$-$\gamma$ mixing
are shown in Fig. \ref{fig:rhogamma}.
These are given by
\be
  \Gamma_{(a+b+c)}^{\rho\gamma} =
  \frac{1+2a_r-a_r^2}{12} (p_\mu p_\nu - p^2 g_\mn)
  \frac{e_rg_r}{{(4\pi)}^2}
  \left[{
    \frac{1}{\bar{\epsilon}} - \ln (-p^2) + \frac{8}{3}
  }\right] ,
\ee
which have {\it no contributions} at $p^2=0$ and hence no counter terms;
\be
  Z_e Z_V^{1/2} Z_a Z_g Z_\pi - 1 = 0 .
\label{Zfactorrhogamma}
\ee
Thus we find that the $\rho$-$\gamma$ mixing also
remains the same as its tree level
in the low energy limit;
\be
  g_\rho(p_\rho^2=0) = a_r g_r f_{\pi r}^2.
\label{rhogammamixpara}
\ee

Comparing Eq.(\ref{Zfactorrhogamma}) with Eq.(\ref{Zfactorrhopipi}),
we have
\be
Z_e = 1 ,
\label{Zefactor}
\ee
which is in accord with the general theorem that
the electromagnetic charge $e$ is not renormalized
by the strong interaction.

{}From Eqs.(\ref{couplinguniversality}) and (\ref{rhogammamixpara}),
we obtain the desired ``low energy theorem" (KSRF I);
\be
  g_\rho(p_\rho^2=0) =
  2 f_{\pi r}^2 g_{\rho\pi\pi}(p_\rho^2=0,p_\pi^2=0,p_\pi^2=0)
\label{KSRFI}
\ee
at one-loop level.

We next investigate loop effects on the
KSRF(II) relation, $m_\rho^2=2f_\pi^2g_{\rho\pi\pi}^2$,
in the low energy limit.
To this end we calculate one-loop graphs for
the $\rho$ meson propagator,
which are shown in Fig. \ref{fig:rhopropagator}.
The contributions to the $\rho$ meson propagator are given by
\be
  \Gamma^\rho_{(a+b+c+d+e)}
  \mathop{\Rightarrow}_{p^2\rightarrow0}
  \frac{3}{2}
  \frac{g_r^2}{{(4\pi)}^2}
  \left({
    \frac{1}{\bar{\epsilon}} - \ln m_\rho^2 + \frac{5}{6}
  }\right) .
\label{gamrhoprop}
\ee
{}From Eq.(\ref{gamrhoprop})
we can determine at $p^2=0$
the counter term
\be
  Z_V Z_a Z_g^2 Z_\pi - 1 =
  - \frac{3}{2}
  \frac{g_r^2}{{(4\pi)}^2}
  \left({
    \frac{1}{\bar{\epsilon}} - \ln m_\rho^2 + \frac{5}{6}
  }\right)
\label{Zfatorrhopropagator}
\ee
in such a way as to obtain the $\rho$ meson mass parameter $M_\rho$
in the low energy limit;
\be
M_\rho^2(p^2=0) = m_\rho^2 = a_r g_r^2 f_{\pi r}^2 .
\label{rhomassparameter}
\ee
Combined with Eq.(\ref{couplinguniversality}), this yields
the KSRF(II) relation for $a_r=2$ in the low energy limit:
\be
  M_\rho^2(p^2=0) =
  2 g_{\rho\pi\pi}^2(p^2=0,k^2=0,q^2=0) f_{\pi r}^2 .
\label{KSRFII}
\ee

Finally, we come to the $\rho$ meson dominance
of the pion electromagnetic form factor.
Let us first determine the counter term for the $\gamma\pi\pi$ vertex;
\be
  - i e_r \epsilon_{3bc} (k-q)_\mu
  \left[{
    Z_e Z_\pi
    \left({
      1-\frac{a_r}{2} Z_a
    }\right)
    -
    \left({
      1-\frac{a_r}{2}
    }\right)
  }\right].
\label{GPPcounterterm}
\ee
We have already determined the renormalization constant $Z_e$
in Eq.(\ref{Zefactor}).
To obtain $Z_a$,
we use the relation
\be
 Z_\sigma = Z_a Z_\pi
\label{ConsistCondSigmaAPi}
\ee
which follows from Eq.(\ref{relFsigFpi}).

$Z_\pi$ and $Z_\sigma$ are determined by
renormalizing the wave functions of the $\pi$ and $\sigma$ fields
at the on-shell point $q^2=0$
(remember that $\sigma$ is massless in the Landau gauge).
The one-loop graphs contributing to these propagators are shown in
Figs. \ref{fig:piprop} and \ref{fig:sigmaprop},
which determine the $\pi$ and $\sigma$
wave function renormalization constants:
\ba
 Z_\pi - 1
&=&
 -
 \left.{
   \frac{d\Gamma_{(a+b+c)}^{\pi\pi}(q^2)}{dq^2}
 }\right|_{q^2=0}
 =
 \frac{3a_r^2}{2} \frac{g_r^2}{{(4\pi)}^2}
 \left[{
   \frac{1}{\bar{\epsilon}} - \ln m_\rho^2 + \frac{5}{6}
 }\right] ,
\label{Zpifactor}
\\
 Z_\sigma - 1
 &=&
 -
 \left.{
   \frac{d\Gamma_{(a+b+c)}^{\sigma\sigma}(q^2)}{dq^2}
 }\right|_{q^2=0}
 =
 \frac{3}{2} \frac{g_r^2}{{(4\pi)}^2}
 \left[{
   \frac{1}{\bar{\epsilon}} - \ln m_\rho^2 + \frac{5}{6}
 }\right] .
\label{Zsigmafactor}
\ea

{}From Eqs.(\ref{ConsistCondSigmaAPi}),
(\ref{Zpifactor}) and (\ref{Zsigmafactor})
we obtain
\be
Z_a-1 = -\frac{3}{2}(a_r^2-1)
   \frac{g_r^2}{{(4\pi)}^2}
   \left[{
     \frac{1}{\bar{\epsilon}} - \ln m_\rho^2 + \frac{5}{6}
   }\right] .
\label{Zafactor}
\ee
We note that for the parameter choice $a_r=1$
the parameter $a$ is not renormalized:
\be
Z_a=1 .
\label{Zaone}
\ee
This implies that in the ``vector limit"\cite{Georgi}
the loop effects of the $\rho$ meson does not induce
deviation from $a=1$.

Combined with Eqs.(\ref{Zefactor}), (\ref{Zpifactor}) and (\ref{Zafactor}),
the counter term for the $\gamma\pi\pi$ vertex
Eq.(\ref{GPPcounterterm}) now reads;
\be
 - i e_r \epsilon_{3bc} {(k-q)}_\mu
 \frac{3a_r(2a_r-1)}{4} \frac{g_r^2}{{(4\pi)}^2}
 \left[{
   \frac{1}{\bar{\epsilon}} - \ln m_\rho^2 + \frac{5}{6}
 }\right] .
\label{gppcounterterm}
\ee

Now, we investigate the one-loop effect on the $\gamma\pi\pi$ vertex.
The graphs which contribute to this vertex are shown
in Figs. \ref{fig:gppdirect} and \ref{fig:gppdirectsub}.
The contributions from each graph
in Fig. \ref{fig:gppdirect} are given by
{
\setcounter{enumi}{\value{equation}}
\addtocounter{enumi}{1}
\setcounter{equation}{0}
\renewcommand{\theequation}{\theenumi.\alph{equation}}
\ba
&{}&
  \Gamma_{(a)}^{\gamma\pi\pi} =
  i e_r \epsilon_{3bc} (k-q)_\mu \frac{a_r}{48}
  \frac{p^2}{{(4\pi f_{\pi r})}^2}
  \left[{
    \frac{1}{\bar{\epsilon}} - \ln (-p^2) + \frac{8}{3}
  }\right] ,
\\
&{}&
  \Gamma_{(b)}^{\gamma\pi\pi} =
  - i e_r \epsilon_{3bc} (k-q)_\mu \frac{1}{24}
  \frac{p^2}{{(4\pi f_{\pi r})}^2}
  \left[{
    \frac{1}{\bar{\epsilon}} - \ln (-p^2) + \frac{8}{3}
  }\right] ,
\\
&{}&
  \Gamma_{(c)}^{\gamma\pi\pi} = 0 ,
\\
&{}&
  \Gamma_{(d)}^{\gamma\pi\pi} =
  i e_r \epsilon_{3bc} (k-q)_\mu \frac{9a_r^2}{8}
  \frac{g_r^2}{{(4\pi)}^2}
  \left[{
    \frac{1}{\bar{\epsilon}} - \ln m_\rho^2 + \frac{5}{6}
  }\right] ,
\\
&{}&
  \Gamma_{(e)}^{\gamma\pi\pi} =
  i e_r \epsilon_{3bc} (k-q)_\mu \frac{3a_r^2}{8}
  \frac{g_r^2}{{(4\pi)}^2}
  \left[{
    \frac{1}{\bar{\epsilon}} - \ln m_\rho^2 + \frac{5}{6}
    + F_{(e)}(p^2)
  }\right] ,
\\
&{}&
  \Gamma_{(f)}^{\gamma\pi\pi} =
  - i e_r \epsilon_{3bc} (k-q)_\mu \frac{3a_r}{4}
  \frac{g_r^2}{{(4\pi)}^2}
  \left[{
    \frac{1}{\bar{\epsilon}} - \ln m_\rho^2 + \frac{5}{6}
    + F_{(e)}(p^2)
  }\right] ,
\\
&{}&
  \Gamma_{(g)}^{\gamma\pi\pi} = 0 ,
\ea
\setcounter{equation}{\value{enumi}}
}
where the function $F_{(e)}(p^2)$ is defined by
\be
F_{(e)}(p^2) \equiv
- \frac{4}{3} \int^1_0 dx \ln\left({1-x\frac{p^2}{m_\rho^2}}\right)
+ \frac{2}{3} \int^1_0 y \, dy \int^1_0 dx
\ln\left({ 1- \frac{xy(1-xy)p^2}{(1-y)m_\rho^2} }\right) .
\ee

Next we calculate the one-loop graphs
through the tree-level direct $\gamma\pi\pi$ vertex
(Fig. \ref{fig:gppdirectsub}).
These contributions are given by
\ba
&{}&
\Gamma^{\gamma\pi\pi}_{(h)} =
  - i e_r \epsilon_{3bc} (k-q)_\mu \frac{a_r^2(2-a_r)}{8}
  \frac{g_r^2}{{(4\pi)}^2}
  F_{(h)}(p^2) ,
\nonum
&{}&
\Gamma^{\gamma\pi\pi}_{(i)} =
  i e_r \epsilon_{3bc} (k-q)_\mu \frac{(2-a_r)(3a_r-4)}{48}
  \frac{p^2}{{(4\pi f_{\pi r})}^2}
  \left[{
    \frac{1}{\bar{\epsilon}} - \ln (-p^2) + \frac{8}{3}
  }\right] ,
\nonum
&{}&
  \Gamma_{(j)}^{\gamma\pi\pi} = 0 ,
\label{gamGPPhij}
\ea
where $F_{(h)}(p^2)$ denotes a certain complicated function
which has no divergent part and $F_{(h)}(p^2=0)=0$.
In the zero photon momentum limit, $p^2=0$,
the contribution of these graphs reduce to
\be
 i e_r \epsilon_{3bc} (k-q)_\mu \frac{3a_r(2a_r-1)}{4}
 \frac{g_r^2}{{(4\pi)}^2}
 \left[{
   \frac{1}{\bar{\epsilon}} - \ln m_\rho^2 + \frac{5}{6}
 }\right] .
\ee

Thus the counter term given in Eq.(\ref{gppcounterterm})
precisely cancels the loop correction at $p^2=0$.
{\it No direct $\gamma\pi\pi$ interaction}
is induced by the one-loop effects
of the hidden gauge bosons {\it in the low energy limit}
for {\it any value of $a_r$}.

However, it is important to investigate the momentum dependence
of the direct $\gamma\pi\pi$ vertex for really
checking the $\rho$ meson dominance.
Actually, away from the zero momentum $p^2=0$, the graphs,
$\Gamma^{\gamma\pi\pi}_{(a)}$, $\Gamma^{\gamma\pi\pi}_{(b)}$,
$\Gamma^{\gamma\pi\pi}_{(e)}$, $\Gamma^{\gamma\pi\pi}_{(f)}$,
$\Gamma^{\gamma\pi\pi}_{(h)}$ and $\Gamma^{\gamma\pi\pi}_{(i)}$,
make contributions to the higher order photon momentum.
Generally, these contributions give rise to the direct $\gamma\pi\pi$ vertex,
thus violating the $\rho$ meson dominance.
 (Of course for $a_r\neq2$, tree level direct $\gamma\pi\pi$ vertex exist
and hence obviously violates the $\rho$ meson dominance.)
However, {\it if we take the parameter choice $a_r=2$},
$\Gamma^{\gamma\pi\pi}_{(a)}$ and $\Gamma^{\gamma\pi\pi}_{(e)}$
are exactly canceled by
$\Gamma^{\gamma\pi\pi}_{(b)}$ and $\Gamma^{\gamma\pi\pi}_{(f)}$, respectively;
\ba
&{}&
\Gamma^{\gamma\pi\pi}_{(a)} + \Gamma^{\gamma\pi\pi}_{(b)} = 0 ,
\nonum
&{}&
\Gamma^{\gamma\pi\pi}_{(e)} + \Gamma^{\gamma\pi\pi}_{(f)} = 0 ,
\label{cancel}
\ea
and $\Gamma^{\gamma\pi\pi}_{(h)}$ and $\Gamma^{\gamma\pi\pi}_{(i)}$
Eq.(\ref{gamGPPhij}) vanish identically,
\ba
&{}&
\Gamma^{\gamma\pi\pi}_{(h)} = 0 ,
\nonum
&{}&
\Gamma^{\gamma\pi\pi}_{(i)} = 0 .
\label{reducezero}
\ea
Therefore {\it no direct $\gamma\pi\pi$ interaction
is induced for all orders of photon momentum,
if and only if we take the parameter choice, $a_r=2$}%
\footnote{For $a_r=2$, Eq.(\ref{reducezero}) as well
Eq.(\ref{cancel}) are correct
not only in the Landau gauge but also in any $R_\xi$ gauge.}.
Incidentally, for $a_r=2$ there is no divergence
in the momentum-dependent part of the
$\gamma\pi\pi$ vertex, namely,
for the electromagnetic form factor of the pion
we need no higher derivative counter term like\cite{GasserLeutwyler}
\be
 L_9
 \Trang{
   F_\mn^{\cal L} \xi_L^{\dag}
    \hat{\alpha}_{\perp}^\mu \hat{\alpha}_{\perp}^\nu \xi_L
  +F_\mn^{\cal R} \xi_R^{\dag}
    \hat{\alpha}_{\perp}^\mu \hat{\alpha}_{\perp}^\nu \xi_R
 } ,
\ee
which is actually needed in the chiral perturbation theory without
hidden gauge boson loop.

Finally, we make some comments on the renormalization-group equations for
the parameters, $a_r$ and $g_r$, in the minimal subtraction scheme.
In this scheme the $\beta$ functions for $a_r$ and $g_r$ are given by
\ba
&{}&
  \beta_a(a_r) \equiv \mu \frac{d a_r}{d\mu}
  = -3 a_r (a_r^2-1) \frac{g_r^2}{{(4\pi)}^2},
\label{betafuncofa}
\\
&{}&
  \beta_g(g_r) \equiv \mu \frac{d g_r}{d\mu}
  = - \frac{87-a_r^2}{12} \frac{g_r^3}{{(4\pi)}^2}.
\label{betafuncofg}
\ea
The $\beta$ function for $a_r$, Eq.(\ref{betafuncofa}), has
an ultraviolet fixed point at $a_r=1$,
which corresponds to the fact that the parameter $a$ is not
renormalized if we set $a=1$ from the beginning (see Eq.(\ref{Zaone})).
Eq.(\ref{betafuncofg}) implies
that the hidden gauge coupling constant $g_r$ is asymptotically free
for not so large value of $a_r$ ($a_r<\sqrt{87}$).
These imply that
for a reasonable value for $a_r$
in the low energy (for example $a_r=2$),
the parameter $a_r$ and the coupling constant $g_r$
go asymptotically to the value of ``vector limit"\cite{Georgi}
($a_r=1$ and $g_r=0$),
i.e., the ``vector limit" is realized
as the ``idealized" high energy limit of the hidden local symmetry.

In conclusion, we have shown that the successful tree level results
of the hidden local symmetry hold at one-loop level.
Thus the predictions of this symmetry has proved
not accidental to the tree level.
In particular, the celebrated KSRF(I) relation survives the loop effects
and hence seems to be a true low energy theorem
as was anticipated in Ref. \cite{BandoKugoYamawakiIII}.
If we further take the parameter choice $a=2$,
the $\rho$-coupling universality and the KSRF(II) relation
also remain valid in the low energy limit.
Most amazingly, the $\rho$ meson dominance still holds at one loop
in the higher photon momentum
not restricted to the low energy limit,
if and only if we take $a=2$.
The ``vector limit" $a=1$ is an ultraviolet fixed point
and is realized as the ``idealized" high energy limit
of the hidden local symmetry Lagrangian.
It is highly desirable to study the full one-loop results of this Lagrangian
at higher momentum $p^2\simeq m_\rho^2$ and see whether or not this
``effective field theory"
survives up to, say, the $A_1$ meson mass region.
Such results are then to be compared
with the generalized hidden local symmetry Lagrangian
having the $\rho$ and $A_1$ mesons on the same footing%
\cite{BandoKugoYamawakiII,BandoFujiwaraYamawaki}.
It would also be worth applying
the full quantum theory of the hidden local symmetry Lagrangian to the
possible vector meson resonances in the
dynamical electroweak symmetry breaking with
large anomalous dimension,
such as the walking technicolor\cite{Holdometc},
the strong ETC technicolor\cite{ETCmodel}
and the top quark condensate models\cite{Miranskyetal}, etc..
Finally, our calculations were done in the Landau gauge for simplicity.
It would be interesting to check our results in other gauges as well%
\footnote{%
A formalism in which the gauge invariance is transparent is
investigated by Tanabashi%
\cite{Tanabashi}}.

We would like to thank Bob Holdom and Masaharu Tanabashi
for stimulating discussion.

\newpage
\def\PR#1#2#3{{ Phys. Rev. }{\bf #1} {(#3)}  #2}
\def\PRL#1#2#3{{ Phys. Rev. Lett. }{\bf #1} {(#3)}  #2}
\def\PL#1#2#3{{ Phys. Lett. }{\bf #1} {(#3)}  #2}
\def\Physica#1#2#3{{ Physica }{\bf #1} {(#3)} #2}
\def\AP#1#2#3{{ Ann. Phys. }{\bf #1} {(#3)}  #2}
\def\ZP#1#2#3{{ Z. Phys. }{\bf #1} {(#3)}  #2}
\def\NP#1#2#3{{ Nucl. Phys. }{\bf #1} {(#3)}  #2}
\def\PREP#1#2#3{{ Phys. Rep. }{\bf #1} {(#3)}  #2}
\def\PROG#1#2#3{{ Prog. Theor. Phys. }{\bf #1} {(#3)}  #2}
\def\SOV#1#2#3{{ Sov. J. Nucl. Phys. }{\bf #1} {(#3)}  #2}
\def\JETP#1#2#3{{JETP}{\bf #1} {(#3)}  #2}
\newcommand{\BandoFujiwaraYamawaki}{%
M.~Bando, T.~Fujiwara and K.~Yamawaki, %
\PROG{79}{1140}{1988}}
\newcommand{\BandoKugoUeharaYamawakiYanagida}{%
M.~Bando, T.~Kugo, S.~Uehara, K.~Yamawaki and T.~Yanagida,
\PRL{54}{1215}{1985}}
\newcommand{\BandoKugoYamawaki}{%
M.~Bando, T.~Kugo and K.~Yamawaki, \PREP{164}{217}{1988}}
\newcommand{\BandoKugoYamawakiII}{%
M.~Bando, T.~Kugo and K.~Yamawaki, %
\NP{B259}{493}{1985}}
\newcommand{\BandoKugoYamawakiIII}{%
M.~Bando, T.~Kugo and K.~Yamawaki, %
\PROG{73}{1541}{1985}}
\newcommand{\CasalbuoniDeCurtisDominiciGatto}{%
R.~Casalbuoni, S.~de~Curtis, D.~Dominici and R.~Gatto, %
\PL{B155}{95}{1985};\NP{B282}{235}{1987}}
\newcommand{\CveticKogerlerII}{%
G.~Cveti\v{c} and R.~K\"ogerler, \NP{B328}{342}{1989}}
\newcommand{\Das}{%
T.~Das, G.S.~Guralnik, V.S.~Mathur, F.E.~Low and J.E.~Young, %
\PRL{18}{759}{1967}}
\newcommand{\EckerGasserLeutwylerPichDerafael}{%
G.~Ecker, J.~Gasser, H.~Leutwyler, A.~Pich and E.~de Rafael, %
\PL{B223}{425}{1989}}
\newcommand{\EckerGasserPichDerafael}{%
G.~Ecker, J.~Gasser, A.~Pich and E.~de Rafael, \NP{B321}{311}{1989}}
\newcommand{\GasserLeutwylerII}{%
J.~Gasser and H.~Leutwyler, \AP{{\rm(N.Y.)}158}{142}{1984}}
\newcommand{\GasserLeutwylerIII}{%
J.~Gasser and H.~Leutwyler, \NP{B250}{465}{1985}}
\newcommand{\Georgi}{%
H.~Georgi, \PRL{63}{1917}{1989}; \NP{B331}{311}{1990}}
\newcommand{\KawarabayashiSuzuki}{%
K.~Kawarabayashi and M.~Suzuki, \PRL{16}{255}{1966}}
\newcommand{\RiazuddinFayyazuddin}{%
Riazuddin and Fayyazuddin, \PR{147}{1071}{1966}}
\newcommand{\Sakurai}{%
J.J.~Sakurai, {\it Currents and Mesons} (Univ. Chicago Press, Chicago, 1969)}
\newcommand{\Wakamatsu}{%
M.~Wakamatsu, \AP{{\rm(N.Y.)}193}{287}{1989}}
\newcommand{\WeinbergCHpT}{%
S.~Weinberg, \Physica{96A}{327}{1979}}

\setlength{\baselineskip}{15pt}

\newpage
\begin{flushleft}
{\large\bf Figure Captions}
\end{flushleft}
\vspace{0.5cm}
\begin{enumerate}
\renewcommand{\labelenumi}{Fig. \theenumi.}
\item 1-particle irreducible graphs contributing %
to the $\rho\pi\pi$ vertex.%
\label{fig:rhopipi}
\item 1-particle irreducible graphs contributing %
to the $\rho$-$\gamma$ mixing.%
\label{fig:rhogamma}
\item 1-particle irreducible graphs contributing %
to the $\rho$ propagator.%
\label{fig:rhopropagator}
\item The 1-loop contributions to the $\pi$ propagator.%
\label{fig:piprop}
\item 1-loop contributions to the $\sigma$ propagator.%
\label{fig:sigmaprop}
\item 1-particle irreducible graphs contributing %
to the $\gamma\pi\pi$ vertex.%
\label{fig:gppdirect}
\item 1-particle irreducible graphs contributing %
to the $\gamma\pi\pi$ vertex %
through the tree level direct $\gamma\pi\pi$ vertex.%
\label{fig:gppdirectsub}
\end{enumerate}


\begin{thebibliography}{99}
\bibitem{BandoKugoYamawaki}
For a review, \BandoKugoYamawaki.   
\bibitem{BKUYY}
\BandoKugoUeharaYamawakiYanagida.
\bibitem{KSRF}
\KawarabayashiSuzuki; \RiazuddinFayyazuddin.
\bibitem{Sakurai}
\Sakurai.
\bibitem{BandoKugoYamawakiII}
\BandoKugoYamawakiII .       
\bibitem{BandoKugoYamawakiIII}
\BandoKugoYamawakiIII.       
\bibitem{Casalbuoni}
\CasalbuoniDeCurtisDominiciGatto.
\bibitem{Weinberg}
\WeinbergCHpT.
\bibitem{GasserLeutwyler}
\GasserLeutwylerII; \NP{B250}{465}{1985}.
\bibitem{Eckeretal}
\EckerGasserPichDerafael; \EckerGasserLeutwylerPichDerafael.
\bibitem{Wakamatsu}
\Wakamatsu; %
W.A.~Bardeen, J.~Bijnens and J.-M.~G\'erard, \PRL{62}{1343}{1989}.
\bibitem{Das}
\Das.
\bibitem{Georgi}
\Georgi.
\bibitem{BandoFujiwaraYamawaki}
\BandoFujiwaraYamawaki; %
K.~Yamawaki, in %
{\it Proc. 1985 INS Symp. on Composite Models of Quarks and Leptons, %
Tokyo, Aug. 13 - 15, 1985}, %
ed. H.~Terazawa and M.~Yasu\`e (INS, Univ. of Tokyo, 1985).
\bibitem{CveticKogerler}
\CveticKogerlerII.
\bibitem{Holdometc}
B.~Holdom, \PL{B150}{301}{1985}; %
K.~Yamawaki, M.~Bando and K.~Matumoto, \PRL{56}{1335}{1986}; %
T.~Akiba and T.~Yanagida, \PL{B169}{432}{1986}; %
T.~Appelquist, D.~Karabali and L.C.R.~Wijewardhana, \PRL{57}{957}{1986}.
\bibitem{ETCmodel}
V.A.~Miransky and K.~Yamawaki, Mod. Phys. Lett. {\bf A4} (1989) 129; %
K.~Matumoto, \PROG{81}{277}{1989}; %
T.~Appelquist, T.~Takeuchi, M.~Einhorn and L.C.R.~Wijewardhana, %
\PL{B220}{223}{1989}.
\bibitem{Miranskyetal}
V.A.~Miransky, M.~Tanabashi and K.~Yamawaki, \PL{B221}{177}{1989}; %
Mod. Phys. Lett. {\bf A4} (1989) 1043; %
Y.~Nambu, Chicago preprint EFI 89-08 (1989); %
W.J.~Marciano, \PRL{62}{2793}{1989}; %
W.A.~Bardeen, C.T.~Hill and M.~Lindner, \PR{D41}{1647}{1990}.
\bibitem{Tanabashi}
M.~Tanabashi, private communication.
\end{thebibliography}
\end{document}